\documentclass[a4paper,fleqn,usenatbib]{mnras}

\usepackage{mathptmx}

\usepackage[T1]{fontenc}
\usepackage{ae,aecompl}

\usepackage{graphicx,color}	
\usepackage{amsmath,mathtools}	
\usepackage{amssymb}	

\newcommand{\bc}{}

\title[The sub-Jovian desert]{Photoevaporation and High-Eccentricity Migration Created the Sub-Jovian Desert}

\author[Owen, J. E. \& Lai, D.]{
James E. Owen$^{1}$\thanks{E-mail: james.owen@imperial.ac.uk} and
Dong Lai$^{2}$
\\
$^{1}$Astrophysics Group, Imperial College London, Prince Consort Road, London SW7 2AZ, UK\\
$^{2}$Department of Astronomy, Center for Astrophysics and Planetary Science, Cornell University, Ithaca, NY 14853, USA
}

\pubyear{2015}

\begin{document}
\label{firstpage}
\pagerange{\pageref{firstpage}--\pageref{lastpage}}
\maketitle

\begin{abstract}
The mass-period or radius-period distribution of close-in exoplanets shows a paucity of intermediate mass/size (sub-Jovian) planets with periods $\lesssim 3$~days. We show that this sub-Jovian desert can be explained by the photoevaporation of highly irradiated sub-Neptunes and the tidal disruption barrier for gas giants undergoing high-eccentricity migration. The distinctive triangular shape of the sub-Jovain desert result from the fact that photoevaporation is more effective closer to the host star, and that in order for a gas giant to tidally circularise closer to the star without tidal disruption it needs to be more massive. Our work indicates that super-Earths/mini-Neptunes and hot-Jupiters had distinctly separate formation channels and arrived at their present locations at different times. 
\end{abstract}




\section{Introduction}

Exoplanet detections have grown at a tremendous rate over the last
decade. The exoplanet population has now reached the point where one
can search the distribution of exoplanet properties for trends which
could elucidate their origins. One of the interesting results to
emerge is the presence of planets with short orbital periods
($\lesssim 10\,$days). There is a well known pile-up of giant planets
with periods around 3~days \citep[e.g.][]{Cumming2008,Howard2010};
these are the ``hot-Jupiters''. Lower mass and smaller planets 
($R_p\sim 1-4~$R$_\oplus$)
have now been found in abundance at short periods as well
\citep[e.g.][]{Borucki2011,Fressin2013,Petigura2013,Silburt2015,Mulders2016}.

However, planets do not populate all regions of parameter space at short periods. 
\citet{Szabo2011} first identified a lack of intermediate mass planets 
($0.02<M_p<0.8~M_J$) at short periods ($P\lesssim 2.5~$days), 
which they labelled the ``sub-Jupiter desert''. 
Working with a planet sample for which the stellar-parameters have been determined accurately using
asteroseismology, \citet{Lundkvist2016} identified a region in the
planet radius-period plane that was devoid of moderate sized 
($2-4$~R$_\oplus$) planets at short periods. 
\citet{Beauge2013} and subsequently \citet{Mazeh2016}
studied the trends in both the planet mass-period and the planet
radius-period planes, showing this hot sub-Jovian desert was present in both data sets.
Furthermore, \citet{Mazeh2016} showed that the desert
consisted of two boundaries: one at high mass/large radius, where the planet's mass/radius decreases with increasing semi-major axis, and another at low mass/small radius, where the planet's mass/radius increases with increasing semi-major axis. 

Until recently, the {\it Kepler} planet candidate catalogue contained
false-positives that made it difficult to study the desert in the
planet radius-period plane. However, using statistical vetting of
planet candidates \citet{Morton2016} removed many of the false
positives. The removal of these false positives showed that the desert
of intermediate sized planets was very clean.  Recently, it has also
been identified that the small (Neptune/sub-Neptune sized) planets
close to the lower boundary are more common around higher metallicity stars \citep{Dong2017,Petigura2018}.
This increased occurrence rate of hot Neptunes around metal-rich stars 
has been interpreted as evidence for high-eccentricity migration  of Neptunes \citep{Dong2017} or metallicity dependent photoevaporation \citep{OMC18}.

Understanding the origin of these features would illuminate the origin
of close-in planets. Several formation mechanisms for these planets have been studied. Many authors
\citep[e.g.][]{Hansen2012,Chatterjee2014,Lee2014,Lee2016,Mohanty2018} 
suggested that low-mass super-Earths/mini-Neptunes formed
{\it in situ}, close to their current orbital locations, while 
\citet{Boley2016,Batygin2016} suggested that hot-Jupiters could also have formed {\it in-situ}. 
On the other hand, disc-driven migration may have played an important
role in the formation of all close-in planets \citep[e.g.][]{Ida2008,Mordasini2009}. For hot Jupiters, an
appealing formation channel is high-eccentricity migration, in which
the planet is pumped into a very eccentric orbit as a result of
gravitational interactions with other planets or with a distant
stellar companion, followed by tidal dissipation which circularises
the planet’s orbit (e.g., Wu \& Murray 2003; Fabrycky \& Tremaine
2007; Nagasawa, Ida \& Bessho 2008; Wu \& Lithwick 2011; Beauge \&
Nesvorny 2012; Naoz et al. 2012; Petrovich 2015; Anderson, Storch \&
Lai 2016; Munoz, Lai \& Liu 2016). However, the origin of the hot sub-jovian desert remains unclear.

\citet{Owen2013} and \citet{Lopez2013} studied the
photoevaporation of low-mass planets that were present at
short-periods at early times and showed that the shape of the
lower-boundary in the radius-period plane was consistent with
photoevaporation of the H/He atmospheres of an initially low-mass 
planets ($\lesssim 20~M_\oplus$). {\bc Additionally, \citet{Jackson2012} and \citet{Kurokawa2014} suggested photoevaporation was important in sculpting the mass distribution of close-in giant planets.}
{\bc Specifically, }\citet{Kurokawa2014} {\bc hypothesised}
that {\bc vigorous} photoevaporation of short-period giant planets {\bc could trigger Roche-lobe overflow}, leaving behind either 
giant planets massive enough to survive {\bc photoevaporation/Roche-lobe} overflow or 
completely photoevaporated solid cores. 
Alternatively, \citet{Matsakos2016} proposed that the entire sub-Jovian desert was created by
tidal disruption of planets in the high-eccentricity migration scenario,
with the upper and lower boundaries of the desert corresponding to different 
mass-radius relation of planets. 

The recent detection of a gap in the planetary radius distribution of
small planets \citep{Fulton2017} confirmed that photoevaporation does
indeed play an important role in the evolution low-mass planet population
\citep{Owen2017,vaneylen2017}. However, detailed
radiation-hydrodynamic photoevaporation models
\citep[e.g.][]{MurrayClay2009,Owen2012,Tripathi2015,Owen2016} indicate
that strong photoevporation of giant planets is difficult, and
removing a giant planet's entire H/He atmosphere is impossible, even
at the shortest orbital periods. 

In this work we explore both photoevaporation and tidal stripping/disruption 
that results from high-eccentricity migration as possible
origins of the sub-Jovian desert. 
Specifically we examine the upper and
lower boundaries in the planet radius - period and mass - period distributions.
We show that a combination of photoevaporation for low-mass planets and tidal disruption 
for massive planets is required to explain the observation.

\section{Planet Sample and Models}

\subsection{Planet Sample}

We chose to represent the location of the planet by its orbital period
rather than semi-major axis. This is because the long-term evolution
of a planet undergoing photoevaporation depends on the total amount of
high-energy flux it receives over its lifetime, which is best
represented by a planet's orbital period when considering a range of
stellar masses \citep{Owen2017} and the period at which a planet is tidally disrupted is independent of stellar mass.

\begin{figure}
\centering
\includegraphics[width=\columnwidth]{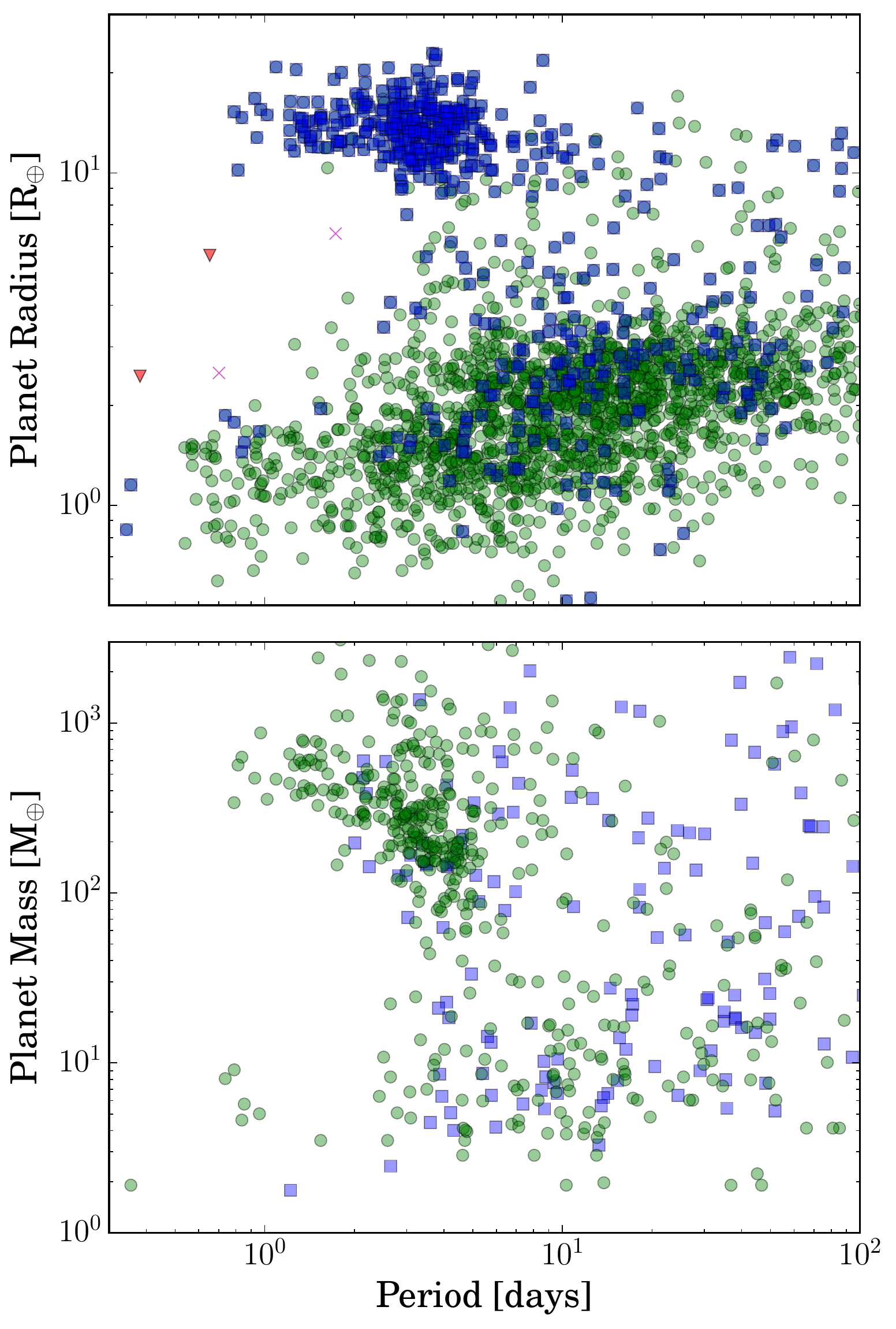}
\caption{Exoplanet radius-period (top panel) and mass-period (bottom panel) distribution. In the radius-period plot the circles are
  confirmed exoplanets, the squares are those with mass measurements,
  the crosses are possible eclipsing binaries identified by
  \citet{Cabrera2017} and the triangles are disintegrating rocky planets
  \citep{Rappaport2012,SanchisOjeda2015}, where the transit radius
  corresponds to the size of the dusty cometary tail escaping the
  planet.  In the
  bottom panel, the squares are $M\sin i$ measurements, whereas the
  circles are mass measurements.}\label{fig:overview}
\end{figure}

We use the confirmed exoplanet database from the NASA planet
archive\footnote{https://exoplanetarchive.ipac.caltech.edu/, downloaded on 3rd August 2017}. We place a wide stellar mass cut on the
sample, retaining those planets whose stellar hosts have masses
between 0.4 and 1.6 M$_\odot$. We show the radius-period and
mass-period exoplanet population in Figure~\ref{fig:overview}. Both plots exhibit a clear sub-jovian desert. {\bc In cases where only $M_p\sin i$ is measured we show this value as the planet's mass (the square symbols in the bottom panel of Figure~\ref{fig:overview}).}
Note that in the radius-period plot (upper panel), two classes of
``misleading'' planets appear in the desert:
(1) The triangles indicate disintegrating rocky planets,
\citep[e.g.][]{Rappaport2012,SanchisOjeda2015}, where the surface
temperatures are large enough to allow sublimation of the planet's surface.
This sublimated rock can then escape the planet's gravity in
a hydrodynamic outflow, before cooling and re-condensing to form dust
\citep{PerezBecker2013}. This dusty outflow then gives the planet a
much larger apparent radius.
(2) The crosses represent planets that appeared in the confirmed exoplanet
database, but have subsequently been suggested to be background
eclipsing binaries \citep{Cabrera2017}.

\subsection{Planetary structure and evolution models}

Throughout this work we will require an understanding of how the
radius of a planet of a given composition varies with both 
its mass and temperature. 
To achieve this we follow \citet{Owen2013,OMenou16,Chen2016} and
use the {\sc mesa} stellar and planetary evolution code
\citep{mesaI,mesaII,mesaIII} to numerically determine the structure
and evolution of a planet. Our planets consist of a solid core
composed of 1/3 iron and 2/3 silicates whose radius is determined
following the mass-radius relationship of \citet{Fortney2007}; this
solid-core is surrounded by a Hydrogen/Helium envelope, with 
the envelope-mass fraction ($X$) given by the ratio of the envelope mass
and the core mass. 
{\bc For each core-mass and envelope mass we select models with initial cooling times in the range 1 to 50 Myr and calculate the evolution of all these models. While the initial cooling can have a small impact on the planet's thermal history \citep[see discussion in][]{Owen2013}, the exact value of the cooling time does not affect the results and conclusions of our work.}
We use the photoevaporative mass-loss rates calculated by
\citet{Owen2012}, following the method set out in \citet{Owen2013}. We
adopt these {\sc mesa} models for all evolutionary
photoevaporative calculations; however, for old massive-planets we use an empirical radius-temperature
relation as described {\bc in Section 2.2.1. In our calculations of lower-mass planets we ignore any possible additional radius inflation mechanism. If an inflation mechanism, such as Ohmic dissipation, operates in lower mass planets it could increase their radii above the values our models find \citep[e.g.][]{Pu2017}, making photoevaporation more effective}.

\subsubsection{Empirical radius-temperature relation for massive planets}\label{sec:inflation}

\begin{figure}
\centering
\includegraphics[width=\columnwidth]{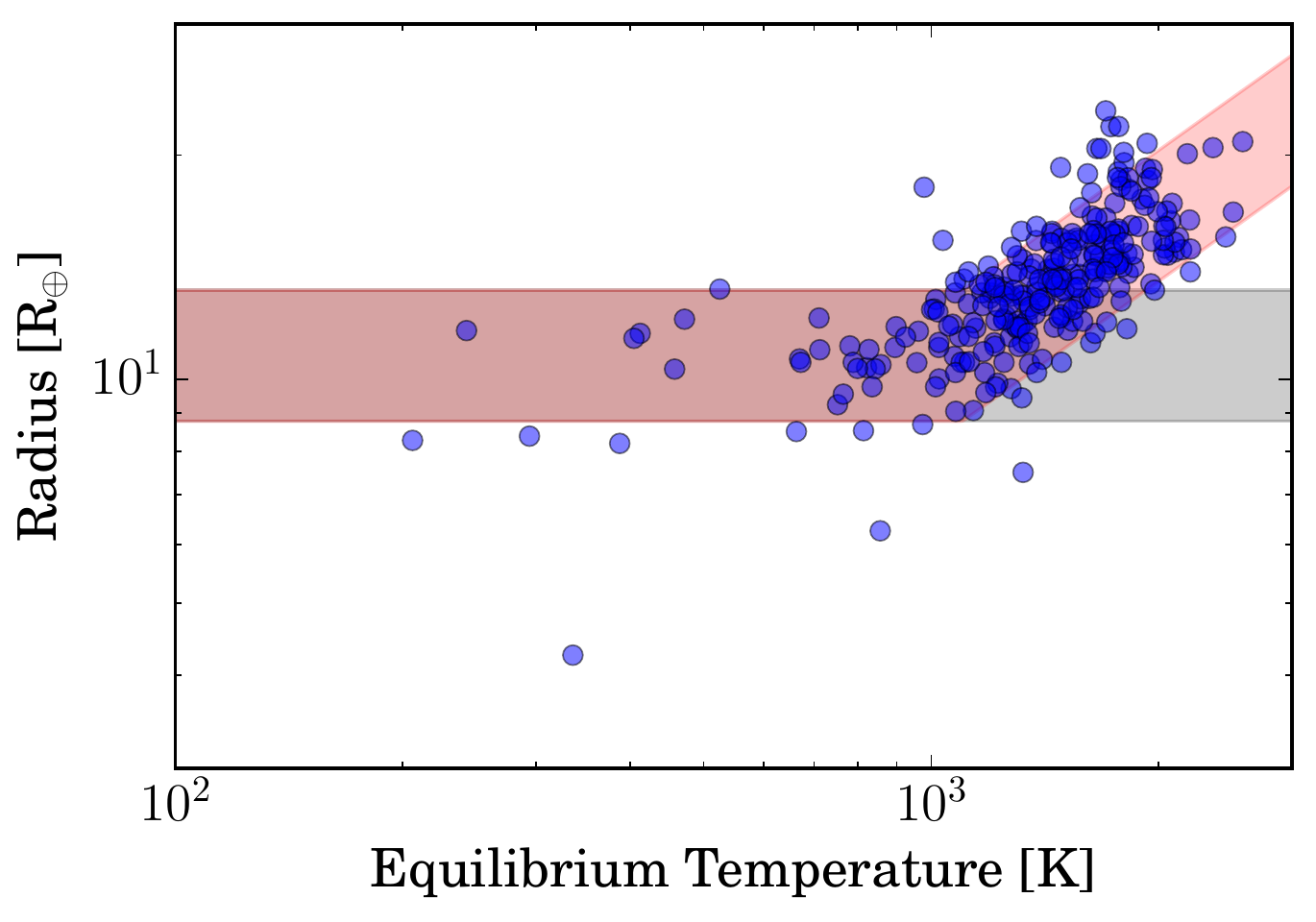}
\caption{Observed exoplanets with masses (or $M_p\sin i$'s) $>$ 0.2
  M$_J$. The inflated radius (Equation~\ref{eqn:inflate}) is shown
  as the red shaded region, and the non-inflated radius is shown as the
  grey shaded region.}\label{fig:empirical}
\end{figure}

Hot Jupiters are known to have inflated radii
\citep[e.g.][]{Baraffe2010,Enoch2012,Thorngren2017}; the origin of this inflation
remains unknown. This means that the mass-radius-period relation cannot
be calculated {\it a priori} theoretically. 
We make use the observed exoplanets to obtain an empirical relation. 
The radius of ``cold'' gas giants is largely independent of mass,
as a result the competition between the degeneracy pressure and Coulomb pressure;
furthermore, hot Jupiter inflation is believed
to be correlated best with the equilibrium temperature ($T_{\rm eq}$,
e.g. \citealt{Laughlin2011,Thorngren2017,Sestovic2018}). 
We therefore adopt the following empirical radius-temperature relation 
that best fits the observations (for planet mass $M_p\gtrsim 0.2M_J$):
\begin{equation}
\frac{R_p}{{\rm R}_J}= f\times
\begin{dcases}
0.8 \left(\frac{T_{\rm eq}}{1100\,{\rm K}}\right)^{0.7224}  & \text{if } T_{\rm eq} > 1100\,{\rm K} \\
0.8 & \text{if } T_{\rm eq} \le 1100\,{\rm K}
\end{dcases}\label{eqn:inflate}
\end{equation}
The factor $f$ is chosen to be between 1 and 1.5, covering the spread in radius at a given equilibrium temperature. This relation is compared to the observed data in Figure~\ref{fig:empirical} (red region). Since inflation may take some time to operate, such that tidally disrupting planets are not inflated when they arrive on their short period orbits, we also chose a non-inflated radius (grey region in Figure~\ref{fig:empirical}) that varies between 0.8 R$_J$ and 1.2 R$_J$. 

\section{Photoevaporation}

\begin{figure}
\centering
\includegraphics[width=\columnwidth]{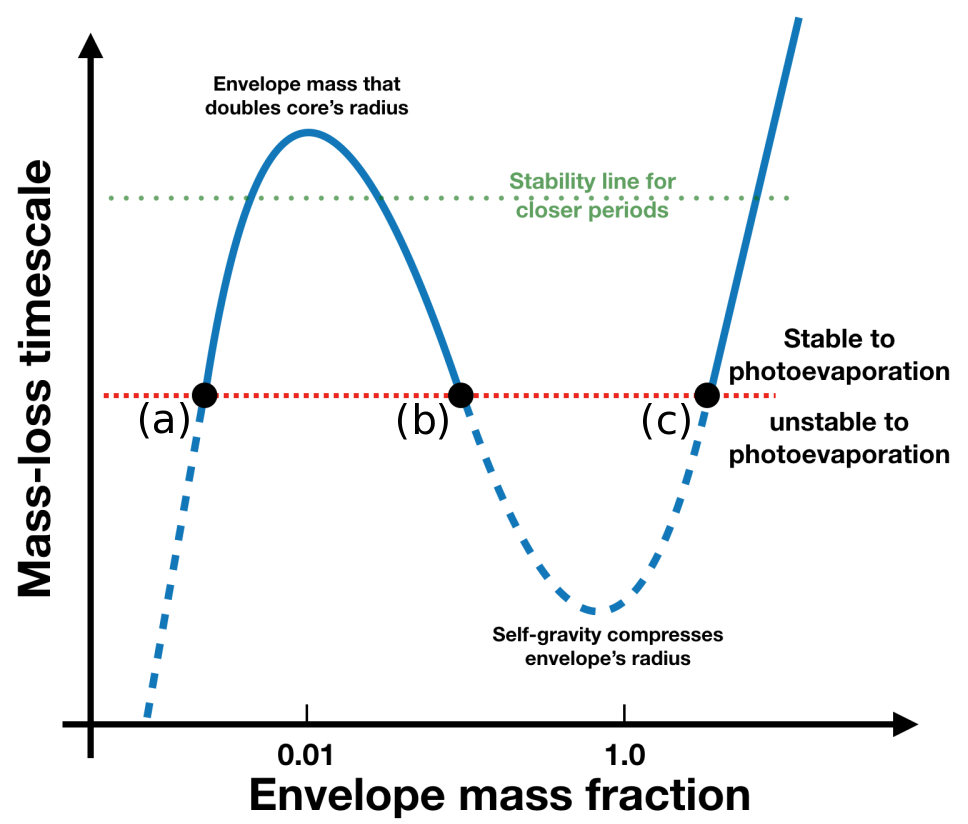}
\caption{Schematic diagram showing the photoevpaorative mass-loss
  timescale (thick blue curve) as a function of the envelope mass
  fraction $X=M_{\rm env}/M_{\rm core}$, {\bc for a fixed core-mass. The horizontal red dashed line denotes 100 Myr, the duration of the bright UV/X-ray phase of the host star. The green dotted line indicates the same 100 Myr line relative to the mass-loss timescale curve, but for shorter-period planets.  The three dots represent three minimally stable atmospheres.}}\label{fig:diagram}
\end{figure}

Exoplanets with H/He atmospheres (envelopes) can lose mass over time through
photoevaporation. Proximity to their parent star results in a planet's
upper atmosphere being heated to temperatures of around
$5,000-10,000$~K by UV/X-ray photons, causing it to escape in a
hydrodynamic wind. Over a planet's lifetime this causes it to lose
mass and {\bc typically} shrink in radius. Since a star is only UV/X-ray bright for of order 100
Myr \citep{Ribas2005,Jackson2012,Tu2015}, photoevaporation
predominately occurs at early times.  \citet{Owen2017} presented a
schematic framework in which to consider the effect of evaporation: 
atmospheres with mass-loss timescale $t_{\dot m}\equiv M_{\rm env}/{\dot m}
\gtrsim 100$~Myr are stable to
mass-loss, while those with $t_{\dot m}\lesssim 100$~Myr are
unstable and evolve towards a lower-mass atmosphere.  For those
atmospheres that are unstable, if a lower-mass state with $t_{\dot m}\gtrsim 100$~Myr exists, then the atmosphere
evaporates to the point where it becomes stable again. If no
lower-mass atmosphere that is stable to evaporation exists then the
planet's atmosphere is completely stripped leaving behind a ``naked'', or ``stripped''
core. The mass-loss timescale of a planet as a function of atmosphere/envelope
mass fraction is schematically shown in Figure~\ref{fig:diagram}
(following \citealt{Owen2017}). 
The curve possess two turning points: The first occurs where the presence
of atmosphere doubles the whole planetary radius (corresponding to 
a H/He atmosphere fraction of $X\sim$ 1\% relative to the core mass);
the second occurs when the atmosphere mass is roughly equal to the core mass,
at which point self-gravity compresses the atmosphere so as to maintain a
a roughly constant planetary radius ($\sim 1-1.5$~R$_J$ for H/He atmospheres);
beyond this second turning point, the mass-loss time increases with the envelope mass. 
Thus, if the first turning point
has $t_{\dot m}\gtrsim 100$~Myr, 
while the second turning point has $t_{\dot m}\lesssim 100$~Myr (as shown by the dotted and
dashed lines in Figure~\ref{fig:diagram}),
then there exists three minimally stable atmospheres (black circles in
Figure~\ref{fig:diagram}): (a) a very low-mass one ($X\lesssim 0.01$) with
the minimum envelope mass required to survive complete stripping,
(b) an intermediate one ($0.01\lesssim X \lesssim 1$) with a maximum
atmosphere mass, and (c) a high-mass ($X\gtrsim1$) one with a minimally
stable atmosphere mass. 
If photoevaportion is the origin of either the
upper or the lower boundary of the unoccupied region in the
radius-period/mass-period plane, then it is the two more massive stable
atmospheres (b \& c) that designate the boundaries (the lowest mass stable atmosphere is the origin of the photoevaporation valley -- \citealt{Owen2017}). Specifically, the low-mass
planet with the maximum atmosphere mass 
stable to photoevpaporation (labelled b)
defines the lower boundary, whereas the high-mass
one with the minimum atmosphere mass (labelled c)
defines the upper boundary.  
At small distance to the star, the mass-loss
timescales of all planets decrease due to the increased high energy
flux. Therefore the stability line, corresponding to a mass-loss
timescale of 100~Myr, crosses the mass-loss curve at a higher point
(represented by going from the red dashed line to the green dotted
line in Figure~\ref{fig:diagram}). Thus, at shorter periods the lower
boundary appears at lower envelope mass fractions (and hence smaller
planetary radii), while the upper boundary due to photoevaporation
appears at higher planet masses.

\subsection{Upper boundary}\label{sec:upper_evap}

\begin{figure}
\centering
\includegraphics[width=\columnwidth]{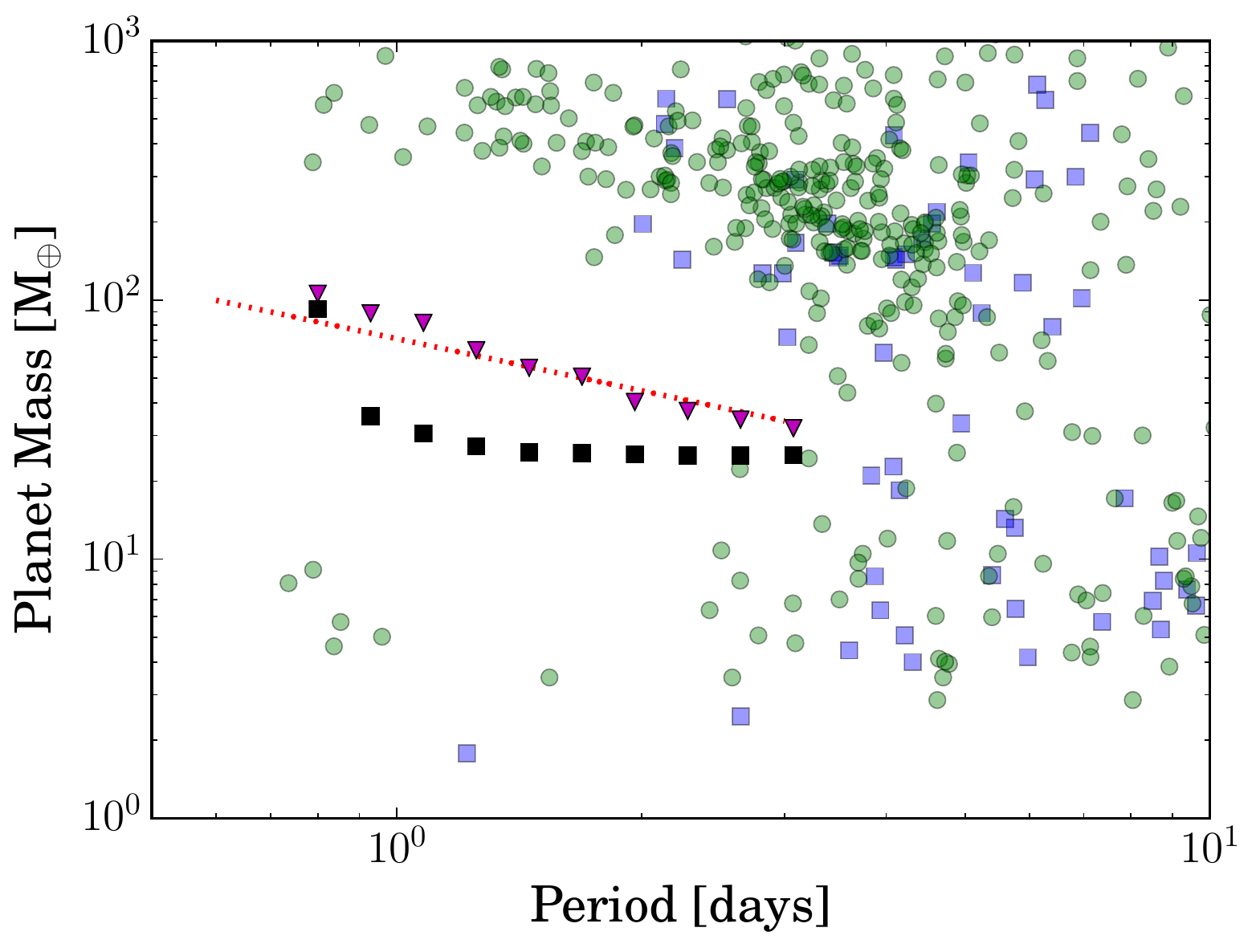}
\caption{Mass-period distribution of close-in exoplanets. The data
  points are the same as in the lower panel of Figure~\ref{fig:overview}. The
  triangles show the initial masses of planets that have envelope mass
  fractions $>1$ after 5 billion years of evolution and the squares
  show the final masses of these planets. The dotted line shows the
  simple $P^{-2/3}$ scaling derived in Section
  ~\ref{sec:upper_evap}.}  \label{fig:evap_failure}
\end{figure}

For the upper boundary due to photoevaporation we take planets to be
massive with an envelope mass fraction 
$X\gtrsim 1$, so the mass-loss
timescale increases with increasing $X$
(see Figure~\ref{fig:diagram}). Before we show the results from numerical
models, we can estimate the result by assuming that the planetary
radii are roughly {\bc constant}. The mass-loss
time-scale $t_{\dot{m}}=M_{\rm env}/\dot{m}$ 
for massive planets with $M_{\rm env}\sim M_p$ is
\begin{equation}
t_{\dot{m}}\propto \frac{a^2M_p^2}{R_p^3L_{\rm HE}},
\end{equation}
where we have crudely used the ``energy-limited'' photoevaporation
mass-loss rate (\citealt{Lammer2003,Baraffe2004}, $\dot{m}\propto
R_{\rm base}^3L_{\rm HE}/(a^2M_p)$, with $L_{HE}$ the high-energy luminosity, $R_{\rm base}$ the radius of the base of the photoevaporative flow and
$a$ the semi-major axis.  
Setting $t_{\dot{m}}= 100$~Myr,
we find that the planet mass on the upper
boundary scales with orbital period ($P$) approximately as $M_p^{\rm
  upper}\propto P^{-2/3}$.  
Looking at Figure~\ref{fig:overview}, it
is clear that the upper envelope empirically begins around a period of
3~days at $\sim 0.1$~M$_J$, so the $P^{-2/3}$ scaling would imply that
even at sub-day periods, planets with sub-jovain masses should be able
to resit photoevaporation. 
This is inconsistent with the observed exoplanet distribution. This inference is confirmed by the full
numerical calculations where we show that the initial and final masses of planets with $X>1$
and a core mass of 10 M$_\oplus$ are able to resist
photoevaporation (see Figure~\ref{fig:evap_failure}).
If photoevaporation were the origin of the upper boundary 
of the sub-jovian desert, the mass-period plane would be filled with sub-jovian mass planets at very short periods. This is clearly not the case, and we must look to another mechanism to explain the upper boundary 
(Section~\ref{sec:high_e}).

Our results are different from \citet{Kurokawa2014}, who argued that
photoevaporation of giant planets can explain the paucity of
sub-jovian planets. There are two reasons for this: Firstly,
\citet{Kurokawa2014} used a combination of energy-limited and
recombination limited photoevaporation models. In the energy-limited
regime they adopted a constant mass-loss efficiency of 25\% following
\citet{Kurokawa2013}, which tends to be a significant overestimate for
giant planets \citep[e.g.][]{Owen2012}, resulting in higher mass-loss
rates. Secondly, in their analysis, the pressure at the base of the photoevpaortive flow is fixed to be 1 nBar; such a choice tends to underestimate the
pressure to which high-energy photons penetrate 
in sub-jovian planets.  In the energy-limited model $\dot{m}\propto R_{\rm base}^3$. Adopting a lower pressure at the base of the flow means $R_{\rm base}$ is larger than in reality and hence the mass-loss rate will be overestimated. {\bc The large mass-loss rates used by \citet{Kurokawa2014} led to mass-loss powered radius inflation, wherein the mass-loss timescale becomes shorter than the cooling time of the planet. When this happens $P{\rm d}V$ work causes the giant planet's envelope to expand, resulting in even larger mass-loss rates and greater envelope expansion \citep[e.g.][]{Baraffe2004}, leading to a runaway. \citet{Kurokawa2014}, indicated this evolutionary pathway would lead to Roche-lobe overflow and catastrophic mass-loss. In our models, due to the lower-mass loss rates, we find mass-loss powered radius inflation of giant planets does not occur and hence catastrophic run-away mass-loss does not happen for close-in hot Jupiters.} This finding is in agreement with work by \citet{Ionov2018}, who also showed massive planets could survive photoevaporation in the desert.

\subsection{Lower boundary}

Having shown that photoevaporation is incapable of explaining the
upper boundary of the sub-jovian desert in the mass-period plane, we now focus on the
lower boundary in both the radius-period and mass-period planes. The
schematic understanding of how photoevaporation creates a
lower-boundary in Figure~\ref{fig:diagram} indicates that these
planets will be of low-mass and have envelope mass fractions in the range from 0.01 to 1. 

In order to investigate the lower-boundary created by photoevaporation
we numerically follow the long-term evolution
of planets with envelope mass fractions initially less than unity as a
function of period. We vary the planet's core-mass and also the
envelope metallicity, where we use the approximate scaling obtained by
\citet{Owen2012} of $\dot{m}\propto Z^{-0.77}$ for the mass-loss
rates, and we assume the bulk metallicity in the entire atmosphere is
constant by scaling the bulk envelope metallicity by the same factor.

\begin{figure}
\centering \includegraphics[width=\columnwidth]{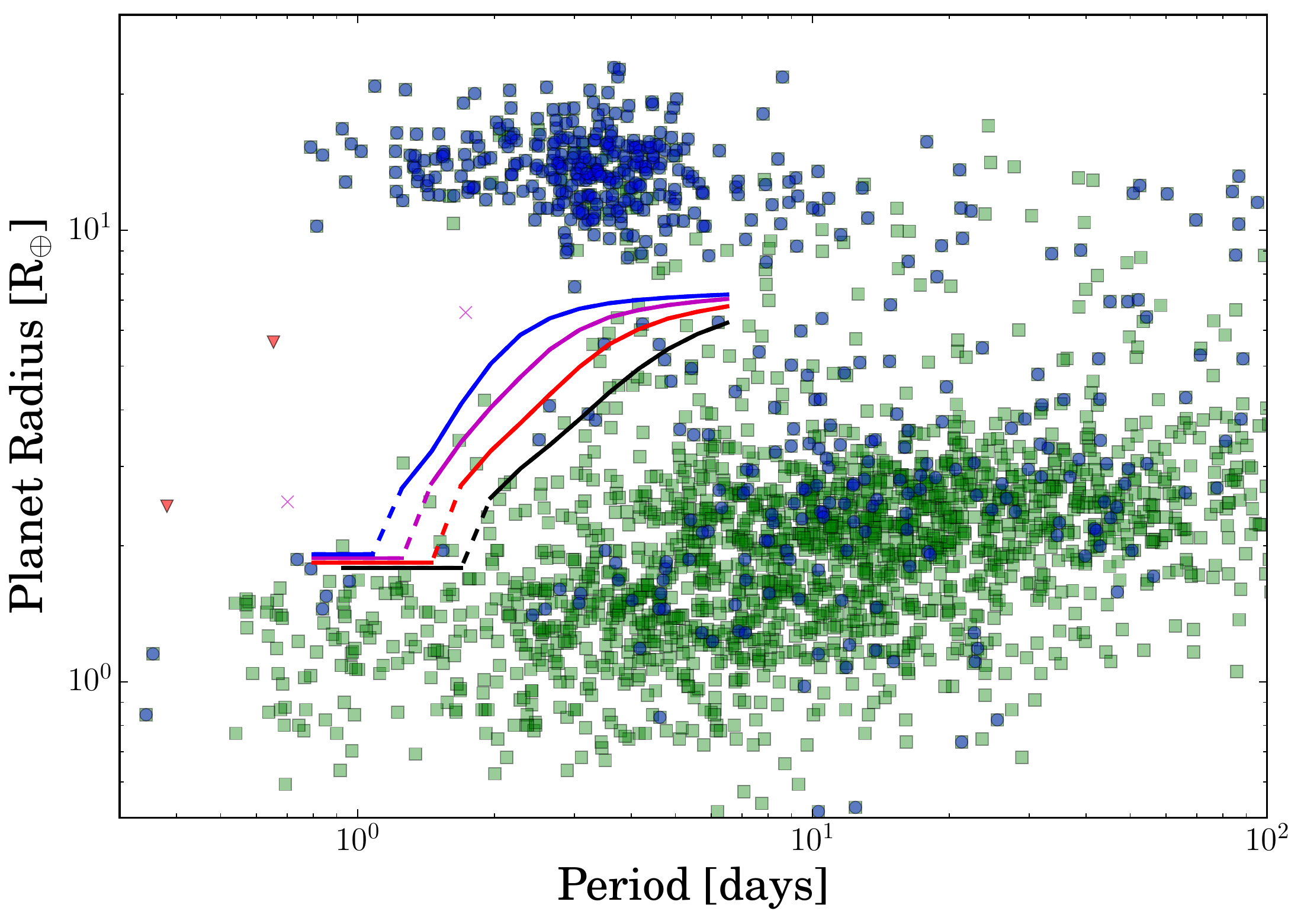}
\caption{Radius-period distribution of close-in exoplanets. The data points 
are identical to those in Figure~\ref{fig:overview}. The lines indicate the boundaries
  determined from the photoevaporation model for different core masses
  (black - 10, red - 11.25, magenta - 12.5 and blue 13.75
  M$_\oplus$). The dashed region of the lines indicate where the
  ``evaporation valley'' \citep{Owen2013,Owen2017} would sit, where
  planets could not exist with stable atmospheres without being
  completely stripped by photoevaporation.}\label{fig:bot_evap_CM}
\end{figure}

\begin{figure}
\centering
\includegraphics[width=\columnwidth]{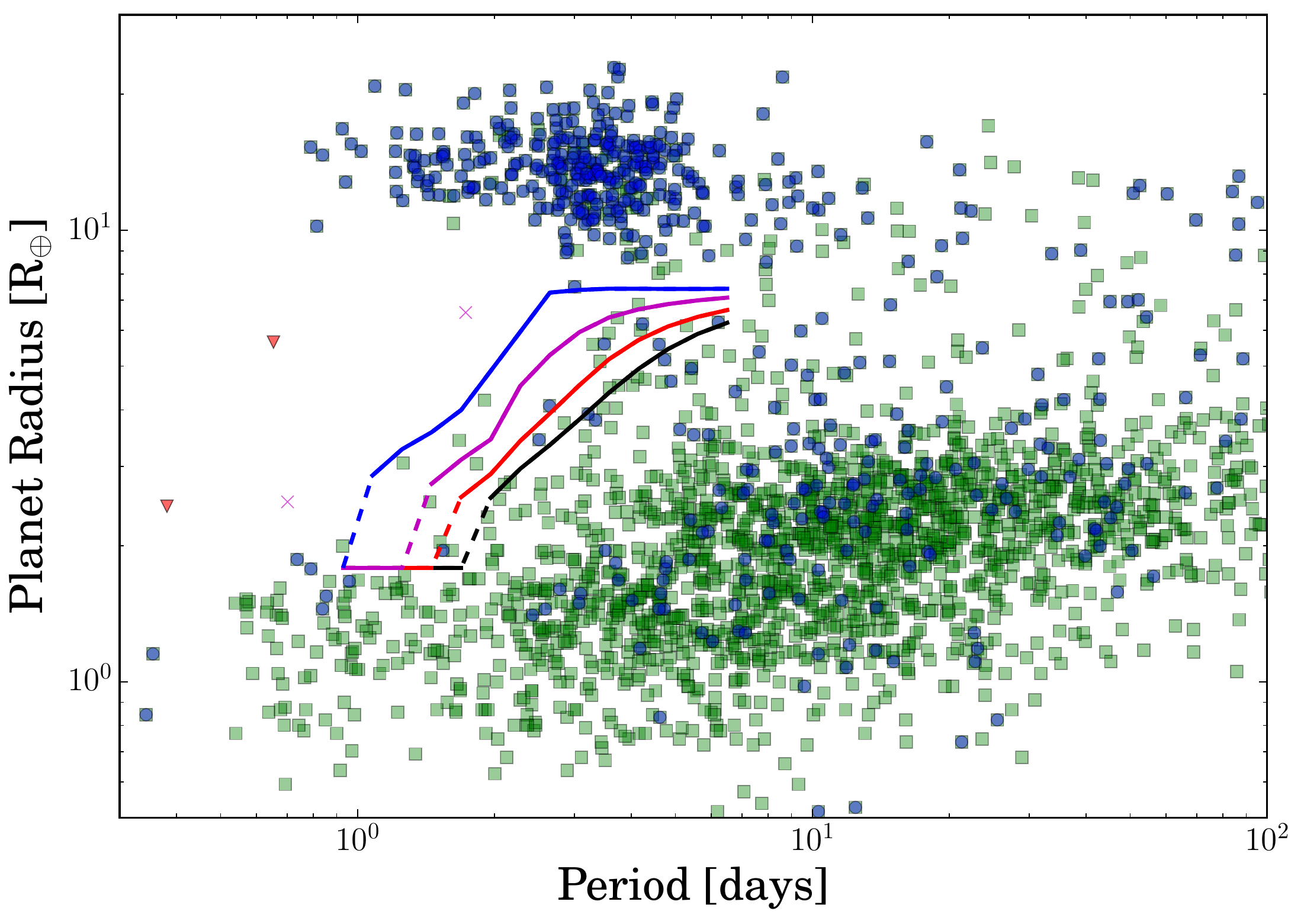}
\caption{Same as Figure~\ref{fig:bot_evap_CM}, but for a fixed core
  mass of 10 M$_\oplus$ and different atmospheric metallicities (black
  -1, red - 1.5, magenta - 3 and blue 10
  Z$_\odot$).}\label{fig:bot_evap_Met}
\end{figure}

We then evolve this initial population of planets under the influence
of photoevaporation for 5 billion years and for each core mass and
atmosphere metallicity and find the largest and most massive planet
that exists at a given period. The results in the radius-period plane
for varying core masses between 10 and 13.75 M$_\oplus$, with solar
metallicity are shown in Figure~\ref{fig:bot_evap_CM} and for
atmospheric metallicity variations in the range 1 to 10 $Z_\odot$,
with a 10 M$_\oplus$ core are shown in
Figure~\ref{fig:bot_evap_Met}. It is clear from these figures that the
shape of the lower boundary in the radius-period plane is well
explained by photoevaporation. 
Comparing the model curves to the data we find the lower-boundary is well explained if the low-mass planet population has a maximum core mass
slightly larger than $\sim 10$~M$_\oplus$, with the exact value
depending on the atmospheric metallicity.
Such a maximum core-mass agrees well with the observed
gap in the radius distribution of small close-in planets
\citep{Fulton2017}, located at 1.8~R$_\oplus$,
corresponding to a stripped solid core of roughly 10~M$_\oplus$.

\begin{figure}
\centering
\includegraphics[width=\columnwidth]{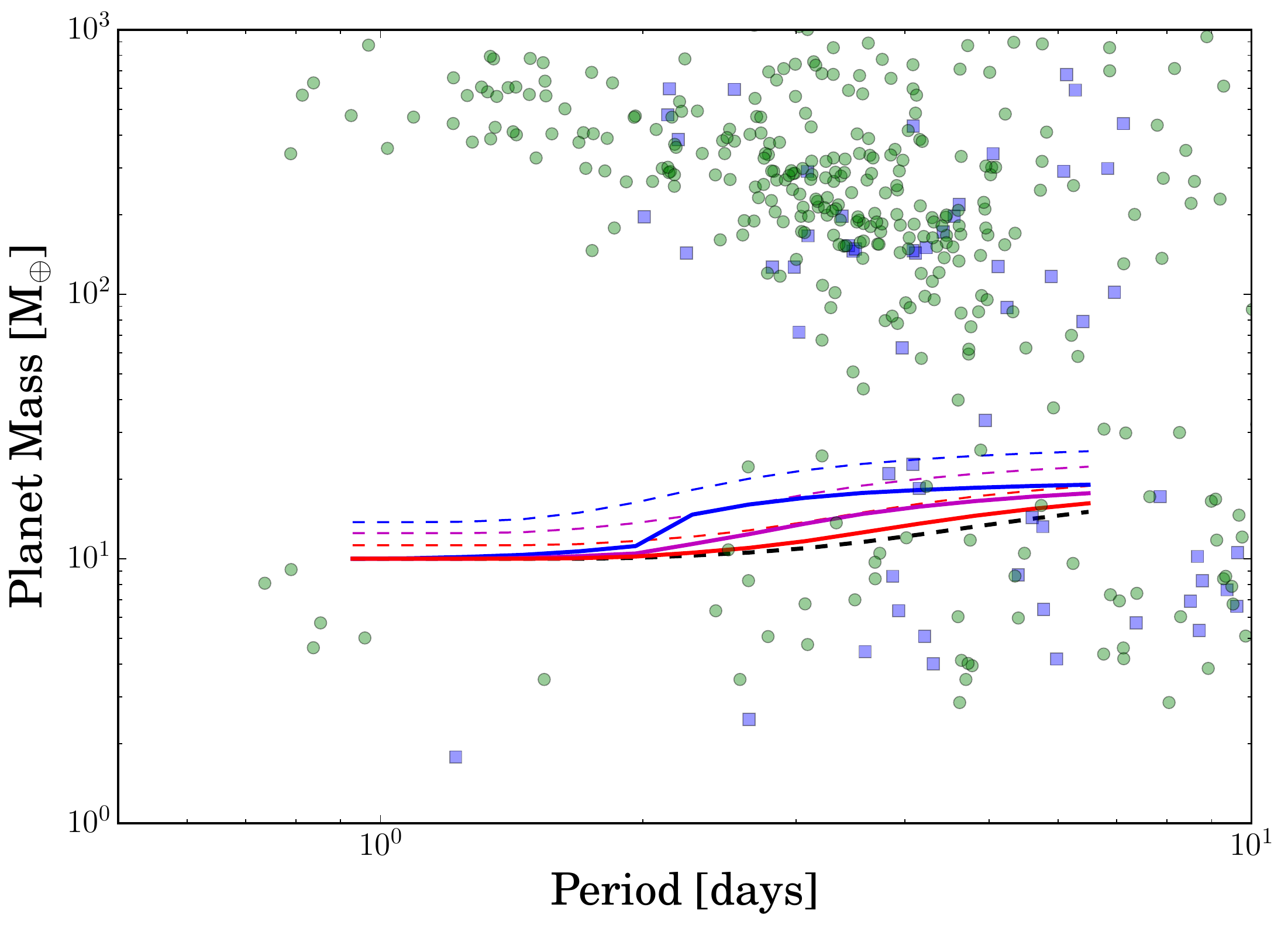}
\caption{Mass-period distribution for close-in exoplanets. The data points
  are the same as shown in the
  lower-panel of Figure~\ref{fig:overview}. The dashed lines represent
  the boundaries obtained from the runs with different core masses and
  are the same as the coloured lines shown in
  Figure~\ref{fig:bot_evap_CM}, the solid lines represent the
  variation from the 10 M$_\oplus$ core due to enhanced metallicities in
  the planet's atmosphere and are the same as the lines shown in
  Figure~\ref{fig:bot_evap_Met}.}\label{fig:lower_evap_mass}.
\end{figure}

As well as investigating the lower boundary in the radius-period
plane, we can compare the model boundaries to the exoplanet data in
the mass-period plane. This comparison is shown in
Figure~\ref{fig:lower_evap_mass}. The boundaries for various
different core masses (dashed lines) and atmospheric metallicities
(solid lines) are shown. The data for small exoplanets with 
measured masses is rather sparse and there is no clear sharp boundary
in the mass-period plane.
All we can say is that the
theoretical boundary in the mass-period plane is consistent with
the data. Finally, checking the few planets that are close to the
theoretical radius-period boundary and have measured masses and radii we
find that they all have masses in the range $10-25$ M$_\oplus$, 
consistent with our expectation that these planets must have a core mass
in the range of $10-15$ M$_\oplus$.

\section{Tidal boundaries and high-eccentricity migration}\label{sec:high_e}


Regardless of the formation channels of close-in planets, their
current (circular) semi-major axes $a$ must  be larger than the critical tidal radius
\begin{equation}
r_{\rm tide}=\eta\left(\frac{M_*}{M_p}\right)^{1/3}R_p,
\end{equation} 
where $\eta=2-3$, depending on the internal structure of the planet.
For concreteness, we adopt $\eta=2.7$ in the following 
based on simulations of giant planet disruption \citep{Guillochon2011}.

In the high-eccentricity migration scenario, a planet is excited into
a highly eccentric orbit due to gravitational interactions with other
planets or with a distant stellar companion, followed by tidal
dissipation in the planet which circularizes the orbit (e.g. Wu \&
Murray 2003; Fabrycky \& Tremaine 2007; Nagasawa, Ida \& Bessho 2008;
Wu \& Lithwick 2011; Beauge \& Nesvorny 2012; Naoz et al. 2012;
Petrovich 2015; Anderson, Storch \& Lai 2016; Munoz, Lai \& Liu
2016). Because tidal circularisation conserves (approximately) the
orbital angular momentum, a planet with a pericenter distance $r_p$
and eccentricity $e\sim 1$ will circularise at a ``final'' semi-major
axis of $a_F=(1+e)r_p\simeq 2r_p$ \citep{Ford2006}. Thus, planets
formed through high-eccentricity migration must satisfy $a_F\ge 2r_{\rm tide}$.

For simple power-law scaling $R_p\propto M_p^\beta P^{-\alpha}$,
the condition $a_F\ge 2r_{\rm tide}$ yields
\begin{equation}
P\ge C_1 \left(\eta^3 M_p^{3\beta-1}\right)^{1/(2+3\alpha)},
\end{equation}
or
\begin{equation}
P\ge C_2 \left(\eta^{3\beta} R_p^{3\beta-1}\right)^{1/(2\beta+\alpha)},
\end{equation}
where $C_1,C_2$ are proportional constants.
In reality, $R_p$ does not have a simple power-law dependence on $M_p$ and $P$
(e.g., For gas giants, the radius 
depends weakly on mass but may depend on the period, while for super-earths 
there is no unique mass-radius relationship as $R_p$ depends on the envelope
fraction). Therefore, for massive planets we use the empirical
radius-temperature relation discussed in Section~\ref{sec:inflation}
and for the small planets we use MESA to calculate the mass-radius
relation for different core masses by setting the age of the planets
to 1~Gyr (our results are fairly insensitive to this choice).

Note that high-eccentricity migration can only form planets with
sufficiently short orbital periods. In order to migrate efficiently, the planet,
starting from an initial semi-major axis $a_0$ ($\sim 1$~au), must be pushed to 
a sufficiently large eccentricity (or sufficiently small pericenter distance $r_p$)
for tidal circularisation to operate. A crude estimate for the orbital decay rate 
in Lidov-Kozai migration is (see Eq.~32 of Anderson et al.~2016, or Eq.~8 of 
Munoz et al.~2016)\footnote{This rate is $(1-e_{\rm max}^2)^{1/2}\simeq (2r_p/a_0)^{1/2}$ 
times the pure tidal decay rate. This factor introduces a small correction to 
Eq.~(\ref{eq:amig}).}
\begin{equation}
\left|{{\dot a}\over a}\right|_{\rm tide,LK}
\simeq 2.8 k_{2p}(\Delta t)_{\rm lag} {GM_\star^2\over  M_p}{R_p^5\over a_0 r_p^7},
\end{equation}
where $k_{2p}$ is the Love number of planet and $(\Delta t)_{\rm lag}$
is the tidal lag time (in the weak tidal friction theory). Requiring 
$|\dot a/a|_{\rm tide,LK}\gtrsim t_{\rm age}^{-1}$, we find that the final (``circularised'')
semi-major axis $a_F\simeq 2r_p$ must satisfy
\begin{eqnarray}
&& a_F\lesssim 0.05\,{\rm au}\, \left({t_{\rm age}\over {\rm Gyr}}\right)^{1/7}
\left({Q_p'\over 10^5}\right)^{-1/7}\left({M_\star\over M_\odot}\right)^{2/7}\nonumber\\
&& \qquad\times \left({a_0\over {\rm au}}\right)^{-1/7}
\left({M_p\over M_J}\right)^{-1/7}
\left({R_p\over R_J}\right)^{5/7},
\label{eq:amig}
\end{eqnarray}
where $Q_p'=Q_p/k_{2p}$, and we have defined 
the tidal quality factor (at the tidal period of 1 day)
via $Q_p^{-1}\equiv (2\pi/{\rm day}) (\Delta t)_{\rm lag}$.

\subsection{Lower boundary}

\begin{figure}
\centering
\includegraphics[width=\columnwidth]{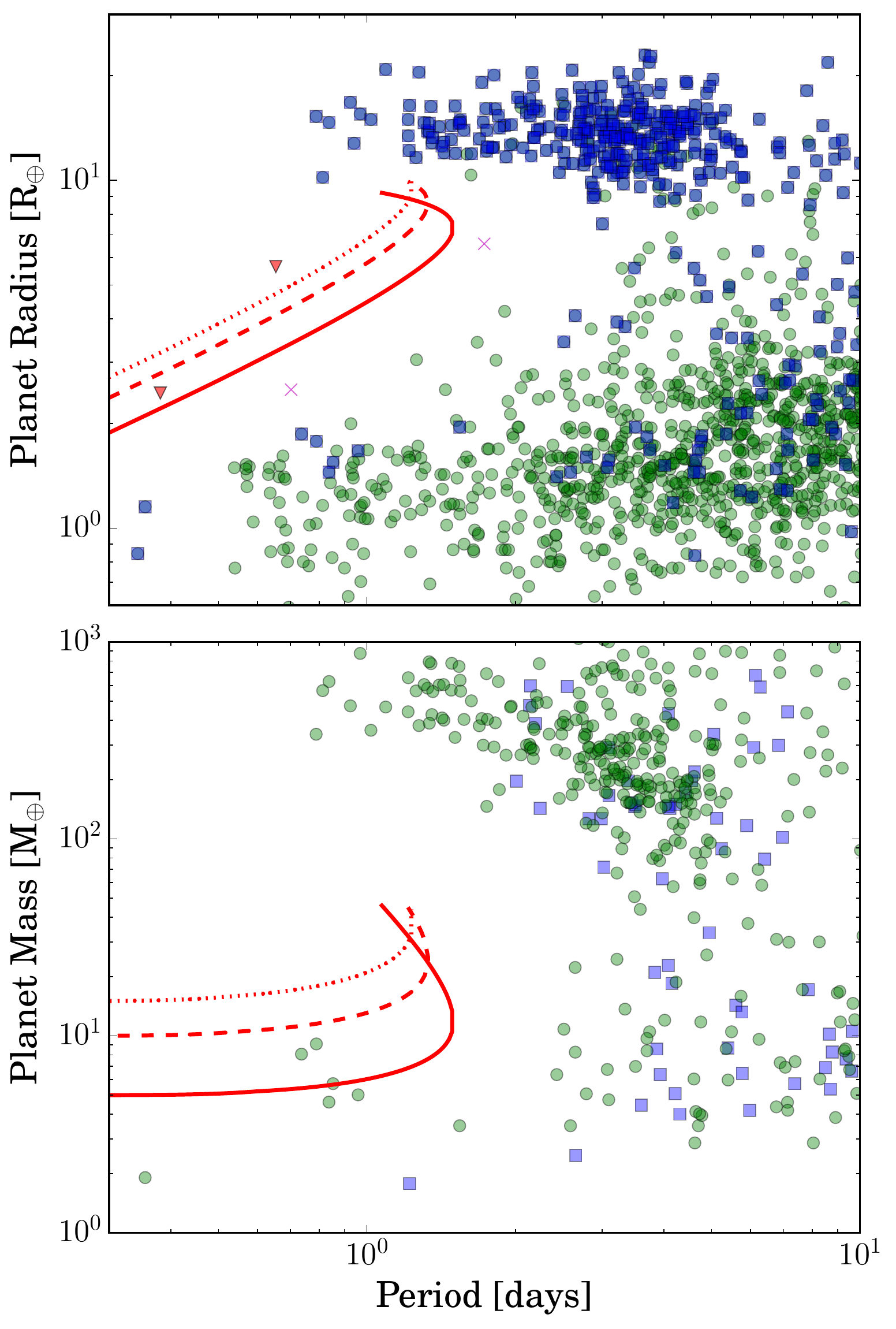}
\caption{Tidal disruption boundary ($a=r_{\rm tide}$) for planets with mass $<0.15$ M$_J$. Each line represents a different core mass with 5 (solid), 10 (dashed) and 15 M$_\oplus$ (dotted) cores shown. In these plots the points represent observed planets and are
  identical to those in Figure~\ref{fig:overview}.}\label{fig:bot_distrubt}
\end{figure}

Figure~\ref{fig:bot_distrubt} shows the tidal disruption boundaries for low-mass
planets ($M_p< 0.15M_J$) in both the radius-period (top panel) and
mass-period (bottom panel) planes.
It is clear that while the observed planets do indeed satisfy the criterion $a>r_{\rm tide}$\footnote{We caution that in making this figure we have assumed a value of $\eta$ which comes from simulated disruptions of giant planets which have different interior structures to the lower mass planets considered here.}, the tidal boundary does not match the paucity of the data. 

\begin{figure}
\centering
\includegraphics[width=\columnwidth]{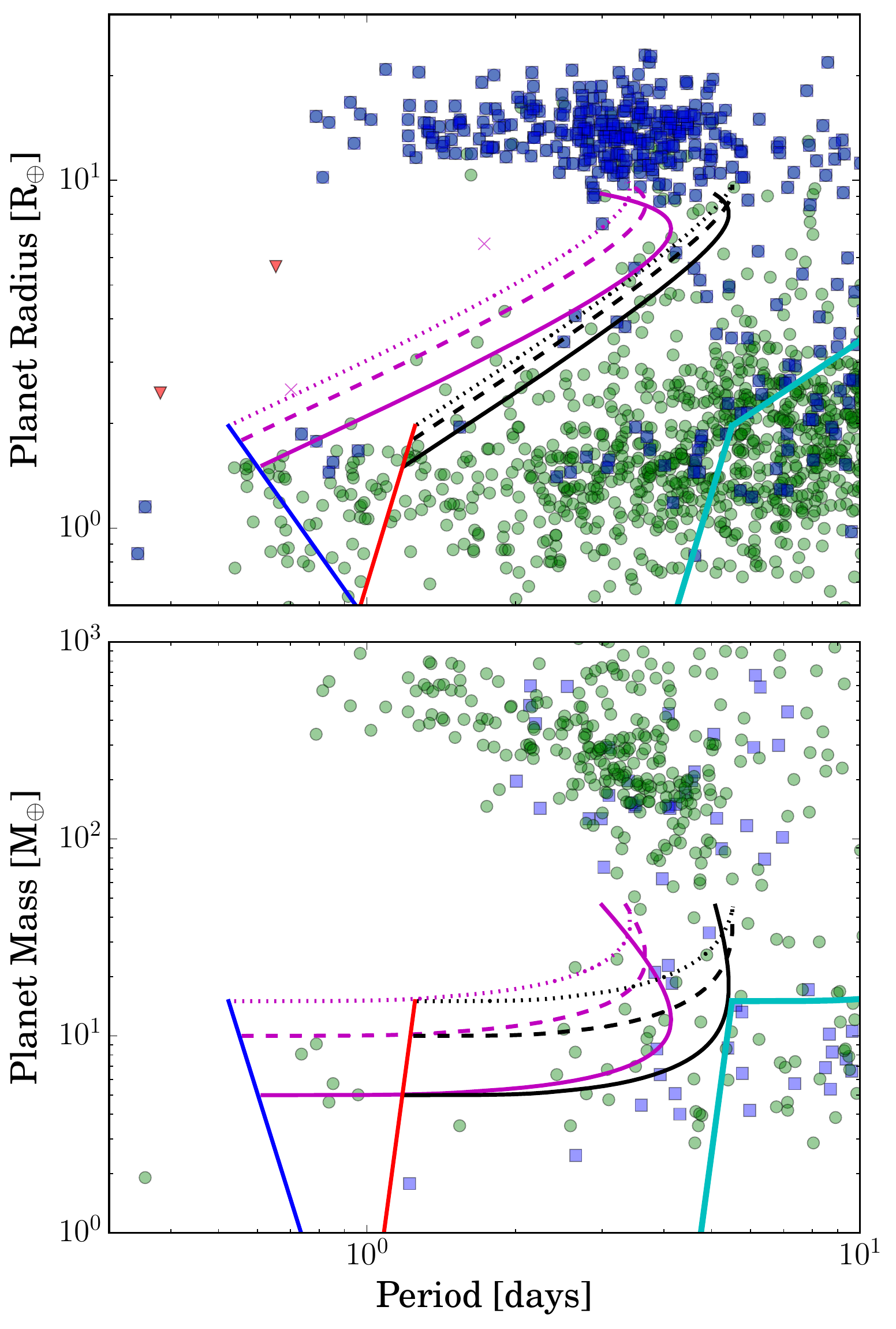}
\caption{The lower boundary of the sub-Jovain desert
in the high-eccentricy migration scenario. The magenta lines correspond to the circularized semi-major axis $a_F=2r_{\rm tide}$. The black lines show the critical periods that planets must attain in order to have circularised during Lidov-Kozai oscillations within 1 Gyr timescale
(Equation~\ref{eq:amig}, with $Q_p'=10^5$).  
The solid blue and red lines show these two boundaries for a solid core with a composition of 1/3 iron, 2/3 silicates and no H/He envelope. The solid, dashed and dotted lines
are the boundaries for planets with H/He atmospheres and core masses
  of 5, 10 and 15 M$_\oplus$ respectively. In order for a planet to have reached its current orbital location via high-eccentricity migration it must reside between the magneta and black lines. The thick solid cyan line shows the same circularisation constraint as the combined constraint of the solid red line and dotted black line, but with $Q_p'=100$ to represent a rocky composition (note that this choice is not appropriate when the planet has a large H/He envelope).  }\label{fig:bot_h_e}
\end{figure}

In Figure~\ref{fig:bot_h_e} we show the boundaries for planets that have undergone high-eccentricity migration (i.e., the circularized semi-major axis $a_F=2r_{\rm tide}$). We also plot the ``circularisation efficiency boundary'' (Equation~\ref{eq:amig}, for $Q'_p=10^5$), indicating that planets that have experienced Lidov-Kozai oscillations must have smaller periods than this boundary in order to circularise on a Gyr or shorter time-scale. {\bc The thick solid cyan line shows the circularisation efficiency constraint for $Q'_p=100$, appropriate for a primarily solid composition, and thus may only be applicable for small, low-mass H/He envelopes.}

The two panels in Figure~\ref{fig:bot_h_e} show that high-eccentricity migration can explain 
{\it parts} of the lower boundary of the sub-jovian desert. However, the range of periods in which planets can reside having formed through a high-eccentricity migration mechanism like Lidov-Kozai oscillations is very narrow and very few planets lie between the tidal disruption boundary (magenta lines) and the circulation efficiency boundary (black lines). Unlike the photoevaporation model these boundaries in the radius-period plane are somewhat insensitive to core mass, with planets with core masses of $\sim$5~M$_\oplus$ allowable close to the boundary. Therefore, the presence of planets with masses $\lesssim 10$~M$_\oplus$ sitting in-between the magenta and black dashed lines would argue that they arrived through high-eccentricity migration. Since high-eccentricity migration is expected to operate on a longer timescale than the 100~Myr timescale of photoevaporation, late-time mass-loss due to photoevaporation is likely to minimal. Therefore we do not expect the high-eccentricity migration lower boundary to be affected by evolution after the planet has circularised. We note that the population of small ($R<2$~R$_\oplus$) planets at sub-day periods could clearly not have been produced by high-energy migration and the bulk of the population of super-earths/mini-neptunes could not have migrated in via Lidov-Kozai oscillations. Therefore, we conclude that while photoevaporation is likely to have sculpted the low-mass planet population creating the lower-boundary it could be polluted by late time high-eccentricity migration of Neptunes/sub-Saturn mass planets.   

\subsection{Upper boundary}

\begin{figure}
\centering
\includegraphics[width=\columnwidth]{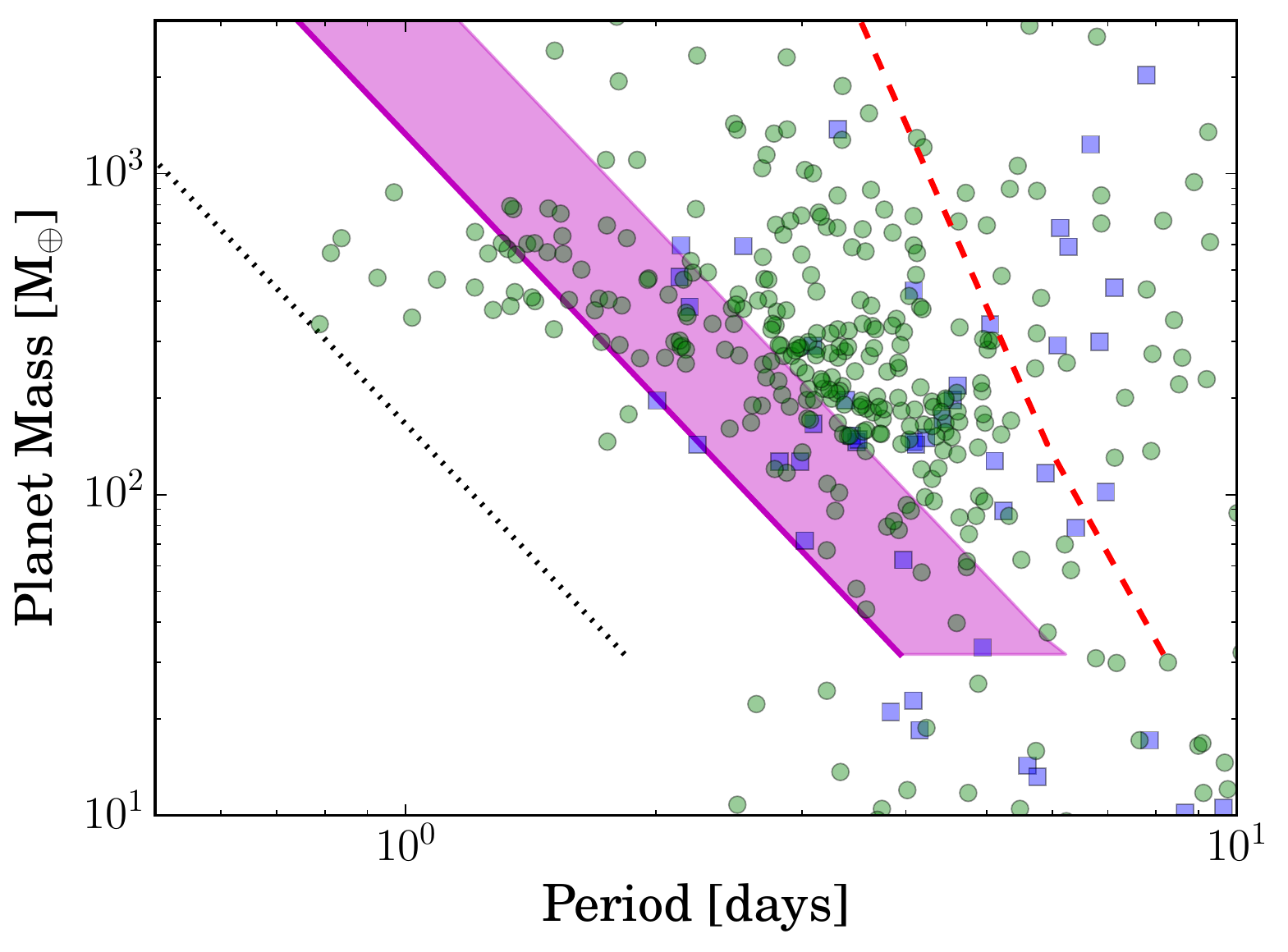}
\caption{The mass-period plane showing the tidal disruption line
  ($a=r_{\rm tide}$; dotted line) and the minimum ciruclarisation period (corresponding to 
  $a=2r_{\rm tide}$; magenta region) and the maximum period for circulations through Lidov-Kozai oscillations on shorter than Gyr timescales (Equation~\ref{eq:amig} with $Q_p'=10^5$; dashed line) 
  for massive planets. 
  The points represent the observed planets and are identical to those in the lower-panel of
  Figure~\ref{fig:overview}. {\bc For clarity, we plot the tidal disruption boundary for its minimum value, while for tidal ciruclarisation we show it as a band, accounting for the spread in the observed radii of giant planets (see section 2.2.1).}}\label{fig:up_envelope_1}
\end{figure}

For giant planets, we use the empirical planet radius relation in
Equation~(\ref{eqn:inflate}). To convert the equilibrium temperature
to orbital period we fix the stellar mass to be 1\,M$_\odot$, stellar
radius to be 1\,R$_\odot$ and stellar effective temperature to
5780~K. In Figure~\ref{fig:up_envelope_1} we show both the tidal boundary 
($a=r_{\rm tide}$, dotted line) the $a=2r_{\rm tide}$ boundary (the magenta band) and the period inside which a planet must reside in order to have circularised on a shorter than Gyr time-scale (Equation~\ref{eq:amig}, dashed line, with $f=1.5$).

As in the case of the lower boundary, the limit $a>r_{\rm tide}$ is satisfied by all planets; however it does not match the observed boundary in the exoplanet population, with ample regions of the planet mass -- orbital period plane where planets could safely reside yet do not. 

\begin{figure}
\centering
\includegraphics[width=\columnwidth]{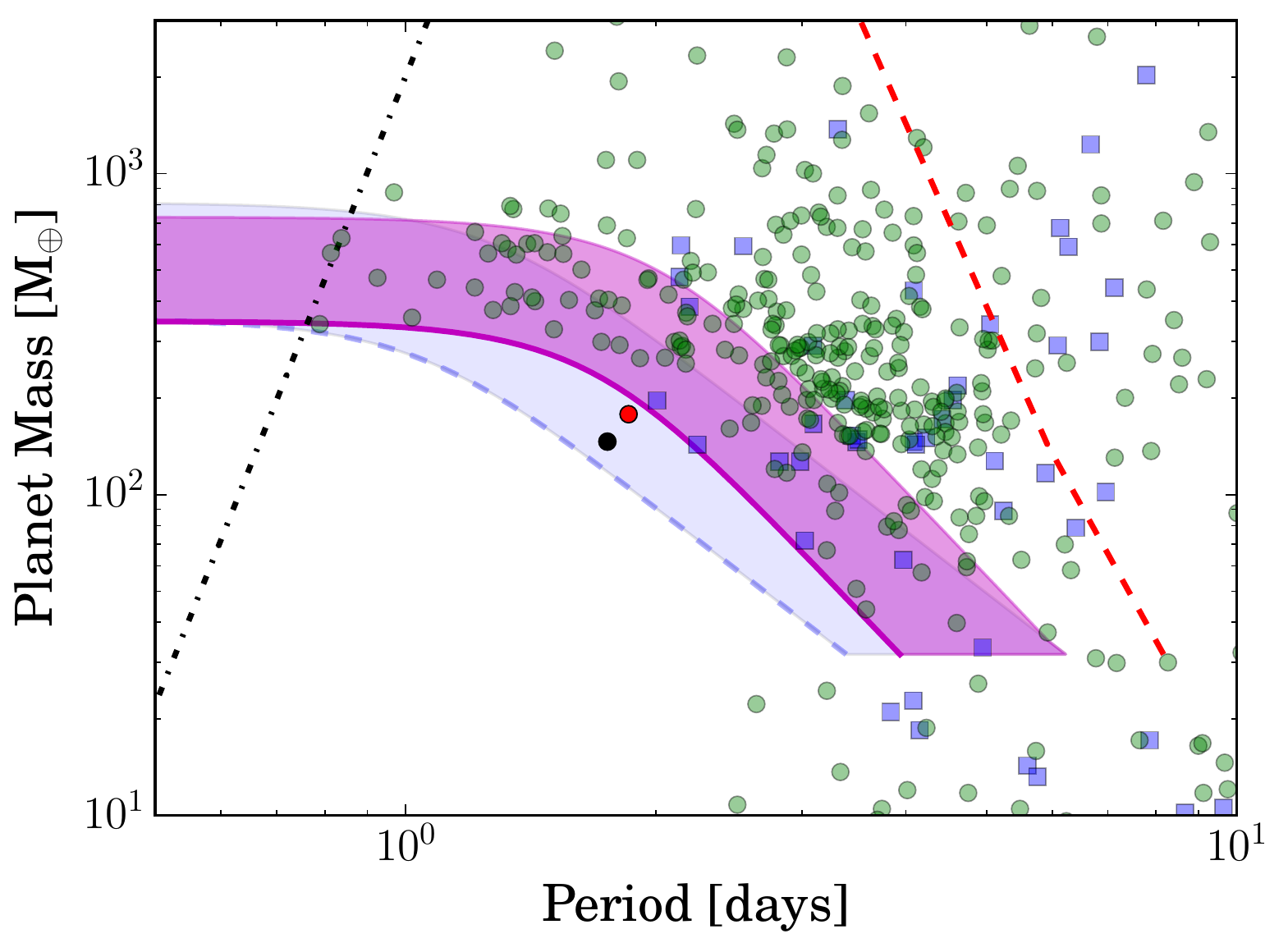}
\caption{Similar to Figure~\ref{fig:up_envelope_1}, but now including
  tidal decay in the high-eccentricity migration boundary. The blue
  region is the same boundary, but arising from planets that were not
  inflated during circularisation. The dot-dashed line corresponds to a
  constant tidal decay time ($\sim 0.1~$Gyr, for a fixed $Q'_*$); planets to the left of
  this line would be undergoing rapid tidal decay and are
  unlikely to be observed. The red point is Kepler-41b and consistent
  with the majenta region within 1$\sigma$. The black point is
  WASP-52b and appears to be an outlier (see
  text).}\label{fig:up_envelope_2}
\end{figure}

Figure~\ref{fig:up_envelope_1} shows that while the circularisation
boundary ($a=2r_{\rm tide}$) provides a reasonable description to the
data, there are a significant number of more massive planets ($\gtrsim
1$~M$_J$) appearing to fill the space between the circularisation
boundary and tidal line ($a=r_{\rm tide}$).
This can be explained by orbital decay due to stellar tides raised by the planet.
The tidal decay timescale decreases with increasing planetary mass, 
therefore Jupiter-mass (and larger) planets 
can be delivered at short periods close to the $a=2r_{\rm tide}$
boundary and subsequently decay in their orbits due to stellar tides.
The orbital decay rate
is given by \citep[e.g.][]{Jackson2008}:
\begin{equation}
\frac{\dot{a}}{a}=-\frac{9}{2}\left(\frac{G}{M_*}\right)^{1/2}\frac{R_*^5M_p}{Q'_*}a^{-13/2},
\label{eqn:tidal_evolve}
\end{equation}
where $Q'_\star$ is the reduced tidal quality factor of the star.  We
integrate Equation~\ref{eqn:tidal_evolve} over a fixed time interval $\Delta t$. 
Since neither $Q'_*$ or $\Delta t$ are known, we constrain the ratio $\Delta t/Q'_*$ by requiring that
the planets in Figure~\ref{fig:up_envelope_1} that reside between the
circularisation limit and the tidal limit reached their current orbits
by tidal decay.  With a choice of $\Delta t/Q'_*$ of 160 years
(corresponding to $Q'_*\sim 10^7$ for $\Delta t\sim 1\,$Gyr), we find
the upper boundary of the sub-jovian desert can be well explained.  
In Figure~\ref{fig:up_envelope_2} we show the resulting
change to the upper boundary in the mass-period plane that results
from tidal circularisation at twice the tidal destruction radius,
followed by tidal decay towards the star (magenta region). 
We also include the boundary (the blue region) 
that would arise if the planets did not become inflated until they had fully circularised; 
in this case we find that the best-fit $\Delta t/Q'_*$ is 20 years. 

In addition, we plot a constant tidal decay time ($\sim 0.1$~Gyr) boundary in
Figure~\ref{fig:up_envelope_2}, shown as the dot-dashed line. 
This boundary arises as one is unlikely to observe planets to the
left of this line as they would be currently undergoing extremely
rapid decay into the star. Two planets stand out as appearing to
be inconsistent with the high-eccentricity boundary: Kepler-41b (red dot) and WASP-52b (black
dot). Kepler-41b is consistent within its
1$\sigma$ mass error bar. WASP-52b is a $\sim 0.5$ M$_J$ mass planet
with a radius of 1.27 R$_J$ \citep{Hebrard2013},
indicating it is not an outlier in terms of giant planet inflation.
WASP-52b does not appear to have any unusually distinguishing features
other than a slightly unusual optical transmission spectrum
\citep{Louden2017}. Either WASP-52b did not reach its present location
by high-eccentricity migration, or it did not become inflated until it
had finished circularising. Both these cases would make WASP-52b an
interesting outlier in the context of hot Jupiter formation.

{\bc As we have fit for the radius-equilibrium temperature relation in Section 2.2.1, our model reproduces the radius-period upper boundary. This match is obviously by construction and not a test. } 

{\bc Our tidal circularisation boundaries are qualitatively similar to those obtained by \citet{Matsakos2016}\footnote{We do not consider their curves associated with 
the ``Secular chaos model'' as defining the locus
 of innermost planets in the period-mass plane: these curves depend on the mass
and semi-major axis of external planet companions,  and thus should be considered
as the necessary conditions for the companion parameters in order to achieve high-eccentricity migration.}; however, we have adopted a more realistic mass-radius-separation relation. This difference results in our model preferring a higher value of $Q_*'$ ($\sim 10^7$ rather than $10^6$); as $Q_*'$ is not constrained theoretically or observationally yet within this range, this difference is not important. Additionally, we have shown (Figure~\ref{fig:bot_h_e}) that only a small region of the lower-mass planets with H/He envelopes could have arrived by high-eccentricity migration, in agreement with population studies of photoevaporation \citep{Owen2017,Wu2018}. }

\section{Implications for the origin of close-in planets}

Understanding the mass/radius vs period distribution of exoplanets
can shed light on the origins of different exoplanet populations.
The presence of the ``evaporation-valley'' in small, close-in exoplanets
\citep{Fulton2017,Owen2017,vaneylen2017} indicates that the majority
of these planets have formed inside the snow-line
and arrived at their current locations before, or soon after the gas disc dispersed. These close-in planets are then sculpted by photoevaporation, producing the lower boundaries in the radius-period/mass-period distributions.

However, it is clear that photoevaporation alone cannot explain both boundaries of the sub-jovian desert, as $M_p\gtrsim 0.5$ M$_J$ planets are able to resist photoevaporation even at extremely short orbital periods. 
If the formation process 
for close-in low-mass planets (that are consistent with the evaporation valley and the lower boundary of the sub-jovian desert) produced a continuum of 
envelope mass fractions $X>1$,
then we should find the planets with intermediate masses at short
periods (i.e. 100~M$_\oplus$ planets in 1 day orbits). Clearly, this is not the case and therefore, whatever the
mechanism that produced the bulk of the low-mass, close-in planet population, it was unable to produce planets with envelope mass fractions 
$\gtrsim 1$. 

For the population of more massive planets ($M_p\gtrsim 0.2$ M$_J$) we find that their distribution in the mass-period plane is consistent with
high-eccentricity migration, with more massive planets surviving closer to their host stars. At masses above $\sim 1\,$M$_J$, tidal decay allows the planets to reach shorter orbital periods after orbital circularisation.
By comparing with the data we require the effective stellar tidal
quality factor to be of order $10^7$, assuming high-eccentricity
migration operates on a Gyr time-scale.  This means that the
majority of more massive planets, with envelope mass fractions $\gtrsim 1$, formed at long periods and then were delivered to their current short-period orbits. The shape of our upper boundary, with tidal decay becoming important above $\sim 1\,$M$_J$, may explain 
the lack of intermediate-mass planets at high irradiation levels, as noted by the recent studies of hot-Jupiter inflation \citep{Thorngren2017,Sestovic2018}.  All giant planets are consistent with this scenario except WASP-52b. 
Therefore, we conclude that the bulk of close-in
giant planets and close-in low-mass planets must have formed through
distinctly different channels at different locations in their nascent
protoplanetary discs and arrived at the short-period orbits on very
different time-scales. 

Our results indicate that {\it in-situ} formation of giant planets either does not happen, or is extremely rare. Furthermore, it seems unlikely that migration of giant-planets through their discs can explain the shape of the upper boundary (particularly the fact that more massive planets can reach shorter periods) as type-II migration is either mass-independent (when the disc mass exceeds the planet mass), or is slower for more massive planets (when the planet mass exceeds the disc mass, \citealt{Syer1995}).
It is of course possible that the planets with masses above 1~M$_J$, for which we have invoked tidal decay to explain their current orbits, could have migrated through discs. Disc migration of such massive giant planets does occur in some planet population synthesis studies \citep[e.g.][]{Ida2008,Bitsch2015}.  While high-eccentricity migration still suffers from several unsolved theoretical problems, it appears to be the dominant process for generating short-period giant planets (see \citealt{Dawson2018} for a recent review).

We note that both photoevaporation and high-eccentricity migration
give similar lower-boundaries in the mass-period and
radius-period planes. The only difference is that high-eccentricity
migration can produce lower mass planets ($\lesssim 5$~M$_\oplus$) near the
radius-period boundary, whereas photoevaporation requires them to be
more massive ($\sim 10$\,M$_\oplus$). However, low-mass planets can only be produced by high-eccentricity migration followed by tidal circularisation in a very narrow range of parameter space, in which very few observed ``hot neptunes'' reside. The current data is too sparse
to test this in any detail. \citet{Dong2017} suggest that these ``hot neptunes'' are mostly singles, and therefore favours high-eccentricity migration.  The metallicity dependence of the boundary, with larger mini-neptunes being common close to their host stars at higher stellar metallicities \citep{Dong2017,Petigura2018}, is also consistent with this suggestion. Due to the narrow range of parameters in which high-eccentricity migration can produce hot neptunes, such metallicity preference should be confined to orbital periods less than a few days. 
\citet{OMC18} show that the uptick in stellar metallicity preference begins at a period of 20~days, and thus is more consistent with a photoevaporative origin.   

We can suggest several possible tests of our scenario. 
The close-in giant planet frequency should be lower around younger stars, as they don't reach their current orbits until late-times. Also, the lower boundary in the radius-period plane should appear at larger radii around younger stars as the planets are still losing mass. 
Alternatively, if a significant fraction of lower-mass planets near the radius-period boundary are produced by high-eccentricity migration, the planetary radius at the boundary should not evolve with time; the boundary should be populated with planets with masses $\lesssim 10$\,M$_\oplus$, and these planets should not have nearby companions and should show evidence for long-period companions. 
Finally, if tracing back in time the evolution of H/He hosting planets near the boundary, accounting for photoevaporative mass-loss (as done for {\it Kepler}-36 b/c in \citealt{OMorton16}) requires an unphysical initial condition (e.g. a planet with an initially large, high-entropy H/He atmosphere that would, in reality, be unbound from the core) then one knows that planet could not have been photoevaporated to its current state.

\section*{Acknowledgements}
We are grateful to the referee for comments which improved the manuscript. JEO is supported by a Royal Society University Research Fellowship.
DL is supported by NASA grant NNX14AP31G and NSF grant AST1715246. JEO is grateful to participants of the 2016 Kavli Summer Program in Astrophysics held at UC Santa Cruz where questions about photoevaporation's efficacy in generating the sub-jovian desert were raised. This research has made use of the NASA Exoplanet Archive, which is operated by the California Institute of Technology, under contract with the National Aeronautics and Space Administration under the Exoplanet Exploration Program.







\bsp	
\label{lastpage}
\end{document}